\documentclass[10pt,nofootinbib,twocolumn]{revtex4-1}

\usepackage[]{graphicx}\graphicspath{{graph/}%
}
\usepackage{amssymb,amsmath}
\newcommand{\p}{{\partial}}
\renewcommand{\vec}[1]{\textnormal{\boldmath$#1$}}

\begin{document}

\title{Generation of a wakefield undulator in plasma with transverse density gradient}\thanks{Work supported in part by the U.S. Department of Energy under contracts No. DE-AC02-76SF00515.}
\author{
G. Stupakov\\ SLAC National Accelerator Laboratory, Menlo Park, CA 94025, USA}
\begin{abstract}
We show that a short relativistic electron beam propagating in a plasma with a density gradient perpendicular to the direction of motion generates a wakefield in which a witness bunch experiences a transverse force. A density gradient oscillating along the beam path would create a periodically varying force---an undulator, with an estimated strength of the equivalent magnetic field more than ten Tesla. This opens an avenue for creation of a high-strength, short-period undulators,  which eventually may lead to all-plasma, free electron lasers where a plasma wakefield acceleration is naturally combined with a plasma undulator in a unifying, compact setup.

\end{abstract}
\maketitle

\emph{Introduction}.---Over the last decade, particle beam-driven plasma wakefield acceleration (PWFA) and laser-driven wakefield acceleration (LWFA) have demonstrated record accelerating gradients and high efficiency in a series of experiments~\cite{Blumenfeld:2007qf,Litos:2014bh,Leemans:2006jt,Steinke:2016mz}. In the long run, they are considered as foundational acceleration techniques for a future multi-TeV scale lepton collider for high-energy physics. On a shorter time scale, there is an intensive effort to employ plasma-assisted acceleration in medical imaging applications, Thomson gamma-ray sources, as well as compact x-ray free electron lasers (FELs)~\cite{Bella_fel}. Radiation  sources based on a combination of LWFA and magnetic undulators have been demonstrated for visible light~\cite{Schlenvoigt:2008rt} and soft x-ray~\cite{Fuchs:2009fr} emission.

It seems very natural to try to combine a plasma-based acceleration with a short-period undulator  field generated inside plasma, in which a relativistic electron bunch would radiate x-rays, making it into a compact, high-brightness, ultrashort x-ray source. Several such concepts have been suggested in the past~\cite{plasma_und_1,plasma_und_2,PhysRevLett.114.145003,Andriyash:2014yq}. They rely on a laser pulse that either excites a plasma oscillation with a transverse to the direction of beam motion component of the field~\cite{plasma_und_1,plasma_und_2,PhysRevLett.114.145003}, or generates the necessary field interacting with periodically positioned nanowires downstream of the acceleration region~\cite{Andriyash:2014yq}. 

In this paper, a different approach to the problem of plasma undulator is proposed. We show that an undulator field can be generated in the plasma wake by simply introducing a \emph{transverse} density gradient into the plasma channel. In our approach, the undulator strength and the period are controlled by the magnitude and the periodicity of the density gradient that oscillates along the path of the beam. Note that the \emph{longitudinal} plasma gradients for control of the beam injection in LWFA has been discussed earlier in the literature~\cite{PhysRevE.58.R5257,PhysRevLett.100.215004,PhysRevLett.86.1011,Gonsalves:2011pd}.

Our analysis is carried out for the PWFA case when the driver is an electron bunch. It can easily be extended for the laser driver by replacing the electromagnetic force from the electron bunch by the ponderomotive force of the laser field~\cite{breizman1992}.

It is well known that plasma is capable to sustain ultrahigh electric fields. The breakdown electric field for the plasma with density $n_0$ is estimated as $E_b\sim ek_p/r_e$, where $k_p=\omega_p/c=\sqrt{4\pi n_0r_e}$ is the inverse plasma skin depth and $r_e = e^2/mc^2$ is the classical electron radius (we use the Gaussian system of units in this paper). When $E_b$ varies on the time scale $\sim \omega_p^{-1}$ and the spatial scale $\sim k_p^{-1}$, it generates the magnetic field $B\sim E_b$. For $n_0 = 10^{17}\ \mathrm{cm}^{-3}$ this estimate gives $E_b \sim 30$ GV/m and $B\sim 100$ T. We expect that in an optimized plasma undulator a considerable fraction of this magnetic field can be used. A solenoidal magnetic field of even much higher strength was demonstrated in computer simulation in the interaction of a screw-shaped laser pulse with under-dense plasma~\cite{Lecz:2016ve}.

\begin{figure}[htb]
\centering
\includegraphics[width=0.3\textwidth, trim=0mm 0mm 0mm 0mm, clip]{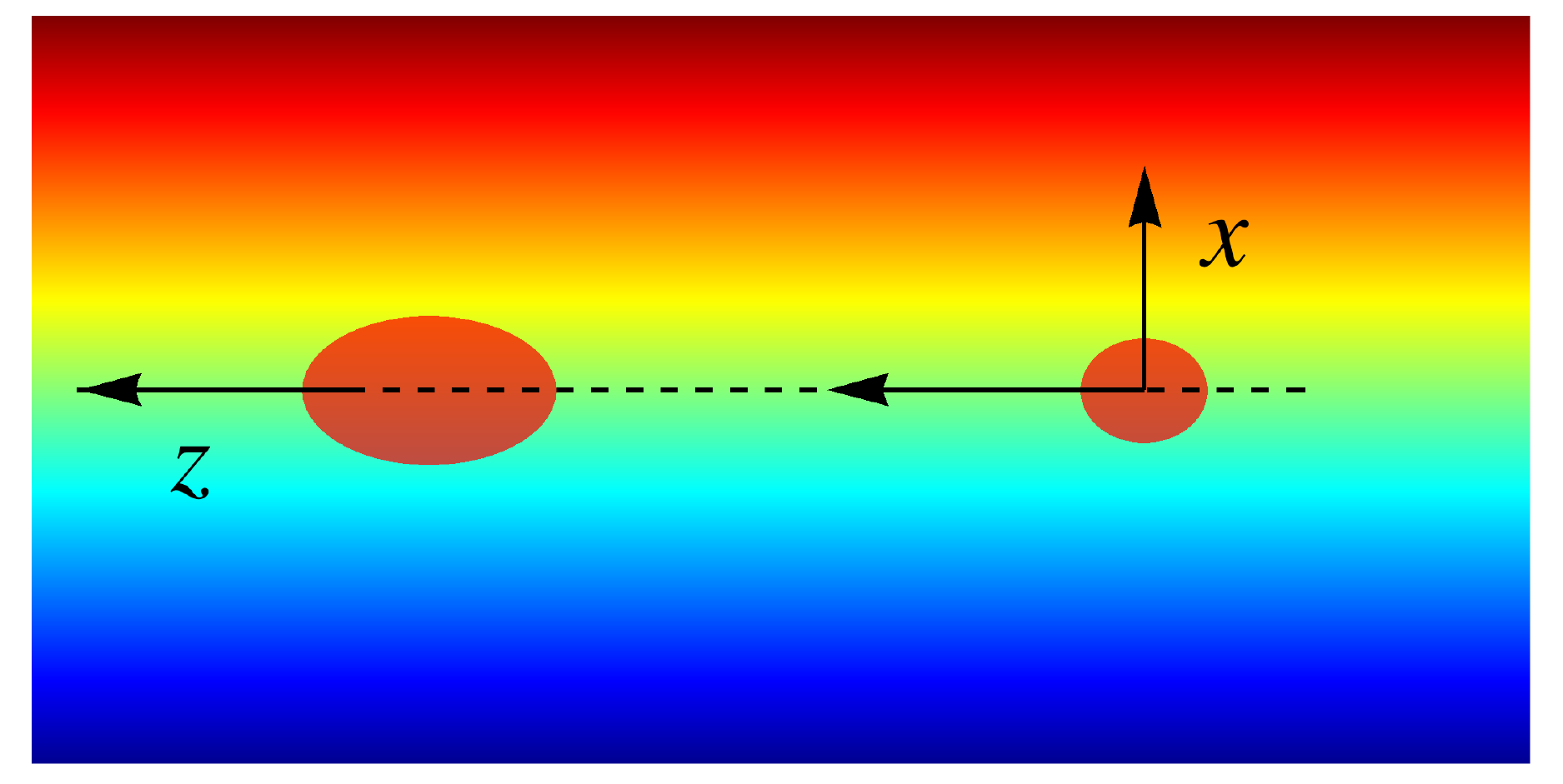}
\caption{A driver beam is followed by a witness bunch propagating in plasma that has a transverse density gradient indicated by the ambient color.}
\label{fig:1}
\end{figure}
\emph{Derivation of the equations for plasma flow}.---We first consider  a plasma that has a transverse density gradient in $x$-direction independent of $z$ and $y$ as shown in Fig.~\ref{fig:1}. The plasma density in the $y-z$-plane is constant and is denoted by $n_0$; the plasma frequency is $\omega_p=ck_p$. Following the standard convention, we normalize time to $\omega_p^{-1}$, length to $c/\omega_p$, and velocities to the speed of light $c$. We also normalize fields to $mc\omega_p/e$, the plasma and beam density to $n_0$, and the current density to $en_0c$. Here $e>0$ is the elementary charge.  

An ultra-relativistic electron driver bunch is moving with $v=c$ along the $z$-axis. The electron beam distribution is assumed axisymmetric, 
    \begin{align}\label{eq:1}
    n_b(r,z,t)
    =
    n_{b0}
    f(r)
    g(\xi)
    ,
    \end{align}
where $\xi = t-z$, $r=\sqrt{x^2+y^2}$, and the functions $f(r)$ and $g(\xi)$ are normalized by unity, $2\pi\int_0^\infty rdrf(r)=1$, and $\int_{-\infty}^\infty g(\xi)d\xi=1$.

To make the problem tractable analytically, we will make two important approximations. First, we will assume that the beam density $n_{b0}$ is small enough that one can treat plasma perturbations and the fields in the plasma in linear approximation, neglecting terms of the second and higher order. We will formulate the second approximation after Eq.~\eqref{eq:11}, when we derive the governing system of equations for a non-uniform plasma.

We consider a steady state in which all fields and perturbations depend on $z$ and $t$ through the  combination $\xi= t-z$. From the expressions for the fields in terms of the scalar potential $\phi$ and vector potential $\vec A$, $\vec E=-\nabla\phi-\p_\xi\vec A$, $\vec B    = \nabla \times \vec A$, it is easy to derive the following relations,
    \begin{align}\label{eq:2}
    \vec E_\perp
    =
    -\nabla_\perp\psi
    -
    \hat{\vec z}\times\vec B_\perp
    ,\qquad
    E_z
    =
    \p_\xi\psi
    ,
    \end{align}
where $\psi = \phi-A_z$, $\nabla = (\p_x,\p_y,-\p_\xi)$, $\hat{\vec z}$ is the unit vector in the $z$-direction, and the subscript $_\perp$ refers to the vector components perpendicular to the $z$ axis. Throughout this paper we use the notation $\p_a$ to denote differentiation with respect to variable $a$.

Note that a relativistic electron moving with $v=c$ along the $z$-axis in the wake of the driver beam experiences the transverse Lorentz force $\vec F_\perp = -\vec E_\perp-\hat{\vec z}\times\vec B_\perp$ which according to the first equation in~\eqref{eq:2} is equal to $\nabla_\perp\psi$. Our goal will be to calculate the transverse gradient of the $\psi$ function at $r=0$ which gives the transverse force \emph{on the $z$ axis}.

In our analysis, the plasma is treated as a cold fluid with ions as immobile, positively charged neutralizing background. The equation of motion for the plasma electrons in linear approximation is
    \begin{align}\label{eq:3}
    \p_\xi
    \vec v
    =
    -
    \vec E
    .
    \end{align}
Using the expression for $E_z$ in Eq.~\eqref{eq:2} and taking into account that both $v_z$ and $E_z$ are equal to zero in front of the bunch, we conclude that
    \begin{align}\label{eq:4}
    v_z
    =
    -\psi
    .
    \end{align}

Denoting the equilibrium plasma density by $n_p(x)$ and the density perturbation by $n_1$, we have  $n=n_p+n_1$, with $|n_1|\ll 1$. With this notation, the equation for the divergence of the electric field reads
    \begin{align}\label{eq:5}
    \nabla\cdot\vec E
    =
    -
    n_1
    -
    n_b
    .
    \end{align}
The beam density $n_b$ is treated together with $n_1$ as a first-order quantity. The density perturbation $n_1$ satisfies the linearized continuity equation,
    \begin{align}\label{eq:6}
    \p_\xi
    n_1
    +
    \nabla
    \cdot
    (n_p\vec v)
    =
    0
    .
    \end{align}
Differentiating this equation with respect to $\xi$, substituting into it Eqs.~\eqref{eq:3}, and using Eq.~\eqref{eq:5} gives an equation that describes an excitation of the plasma oscillations in a nonuniform plasma by the beam,
    \begin{align}\label{eq:7}
    \p_{\xi\xi}
    n_1
    +
    n_p
    n_1
    =
    -n_pn_b
    +
    \vec E_\perp
    \cdot
    (\nabla_\perp
    n_p)
    .
    \end{align}

We now derive equations for $\psi$ and $\vec B_\perp$. Substituting Eq.~\eqref{eq:2} into Eq.~\eqref{eq:5} we obtain
    \begin{align}\label{eq:8}
    \nabla_\perp
    \cdot
    (
    \nabla_\perp\psi
    +
    \hat{\vec z}\times\vec B_\perp
    )
    +
    \p_{\xi}
    E_z
    =
    n_1
    +
    n_b
    .
    \end{align}
Using then the Maxwell equation $\nabla\times \vec B = \p_\xi \vec E + \vec j$ with the current density
    \begin{align}\label{eq:9}
    \vec j = -n_p\vec v-n_b\hat{\vec z}
    ,
    \end{align}
we eliminate the magnetic field from Eq.~\eqref{eq:8} and arrive at $\Delta_\perp\psi=n_1-n_pv_z$ which, with account of relation~\eqref{eq:4}, can be written as
    \begin{align}\label{eq:10}
    \Delta_\perp\psi
    -
    n_p\psi
    =
    n_1
    .
    \end{align}
Here $\Delta_\perp$ is the transverse part of the Laplacian operator. To obtain an equation for $\vec B_\perp$ we take the curl of both sides of the Maxwell equations $\nabla\times \vec B=\p_\xi \vec E+\vec j$ which gives $\Delta_\perp\vec B = -\nabla\times \vec j$. Taking the transverse part of this equation yields
    \begin{align}\label{eq:11}
    \Delta_\perp\vec B_\perp
    = 
    \hat{\vec z}
    \times
    \nabla_\perp j_z
    +
    \hat{\vec z}
    \times 
    \p_\xi\vec j_\perp
    .
    \end{align}
Eqs.~\eqref{eq:7}, \eqref{eq:10} and~\eqref{eq:11}, with relations~\eqref{eq:2} and~\eqref{eq:3} and the connection between the current density and the velocity~\eqref{eq:9} constitute a full set of equations for the three unknowns $n_1$, $\psi$, and $\vec B_\perp$.

\emph{Solving equations in perturbation theory}.---For practical purposes, the equilibrium dimensionless plasma density is approximated by the linear profile,
    \begin{align}\label{eq:12}
    n_p(x) = 1+\alpha x,
    \end{align}
where $\alpha$ is the dimensionless density gradient. We will now formulate our second approximation: the parameter $\alpha$ is treated as small, and we will solve our equations using  perturbation theory that neglects terms of the second and higher order in $\alpha$. In physical units, our assumption means that the plasma density changes in $x$ direction on the scale much larger than $k_p^{-1}$.

In the zeroth-order approximation in $\alpha$ corresponding to setting $\alpha=0$, we have $n_p=1$ which means a uniform plasma. This is the problem formulated and solved in the original pioneering papers on  PWFA~\cite{PhysRevLett.54.693,chen_etal}. Here we will reproduce the main elements of their solution needed for our next approximation. As is clear from the axisymmetry of the problem, in this approximation, there is no transverse force acting on a relativistic particle on the axis of the system. The magnetic field has the only component $B_\theta$ that together with $n_1$ and $\psi$ depend on $r$ and $\xi$. We will use the superscript $^{(0)}$ to mark the variables in this order. The equations that define these three quantities are:
    \begin{align}\label{eq:13}
    &\p_{\xi\xi}
    n_1^{(0)}
    +
    n_1^{(0)}
    =
    -n_b
    ,\qquad
    r^{-1}\p_r (r \p_r\psi^{(0)})
    -
    \psi^{(0)}
    =
    n_1
    ,\nonumber\\
    &\p_r(r^{-1}\p_r rB_\theta^{(0)})
    -
    B_\theta^{(0)}
    =
    -
    \p_rn_b
    .
    \end{align}
For the beam distribution function~\eqref{eq:1}, their solution is given by the following equations,
    \begin{align}\label{eq:14}
    n_1^{(0)}
    &=
    n_{b0}
    f(r)
    G(\xi)
    ,\qquad
    \psi^{(0)}(r,\xi)
    =
    n_{b0}
    F_1(r)
    G(\xi)
    ,\nonumber\\
    B_\theta^{(0)}(r,\xi)
    &=
    n_{b0}
    F_2(r)
    f(\xi)
    ,
    \end{align}
where the functions $G$, $F_1$ and $F_2$ are defined by the relations
    \begin{align}\label{eq:15}
    G(\xi)
    &=
    -
    \int_{-\infty}^\xi
    d\xi'
    g(\xi')
    \sin(\xi-\xi')
    ,\nonumber\\
    F_1(r)
    &=
    -
    K_0(r)
    \int_0^r
    r'dr'
    I_0(r')
    f(r')
    ,\nonumber\\
    &-
    I_0(r)
    \int_r^\infty
    r'dr'
    K_0(r')
    f(r')
    ,\nonumber\\
    F_2(r)
    &=
    K_1(r)
    \int_0^r
    r'dr'
    I_1(r')
    f'(r')
    ,\nonumber\\
    &+
    I_1(r)
    \int_r^\infty
    r'dr'
    K_1(r')
    f'(r')
    ,
    \end{align}
where $I_0$, $I_1$, $K_0$ and $K_1$ are the modified Bessel functions of the first and second type  of order zero and one, respectively.

In the next, linear approximation in $\alpha$, we represent the density perturbation $n_1$ and $\psi$ as a sum of the zero-order terms found above and corrections $\delta n_1$ and $\delta \psi$ due to the nonuniformity of the plasma density:
    \begin{align}\label{eq:16}
    n_1
    =
    n_1^{(0)}
    +
    \delta n_1
    ,\qquad
    \psi
    =
    \psi^{(0)}
    +
    \delta\psi
    .
    \end{align}
Substituting these relations into Eqs.~\eqref{eq:7} and~\eqref{eq:10} and using Eq.~\eqref{eq:12} for $n_p$ we obtain
    \begin{align}\label{eq:17}
    \p_{\xi\xi}
    \delta n_1
    +
    \delta n_1
    =
    \alpha
    (
    -
    r
    n_1^{(0)}
    -
    \p_r\psi^{(0)}+B_\theta^{(0)}
    )
    \cos\theta
    ,
    \end{align}
and
    \begin{align}\label{eq:18}
    \Delta_\perp\delta \psi
    -
    \delta \psi
    =
    \alpha
    (
    \delta n_1
    +
    r\psi^{(0)}\cos\theta
    )
    ,
    \end{align}
where we have used the relation $E_r^{(0)}=-\p_r\psi^{(0)}+B_\theta^{(0)}$ that follows from Eq.~\eqref{eq:2} in the zeroth order. An inspection shows that in the cylindrical coordinate system $r,\theta,z$ both $\delta n_1$ and $\delta \psi$ have an angular dependence $\propto\cos\theta$ and can be written as
    \begin{align}\label{eq:19}
    \delta n_1
    =
    u(r)\cos\theta
    ,\qquad
    \delta\psi
    =
    w(r)\cos\theta
    .
    \end{align}
The functions $u$ and $w$ are found from Eqs.~\eqref{eq:17} and~\eqref{eq:18}; as pointed out above, we are interested in the on-axis force $\vec F_\perp = \nabla_\perp\delta\psi|_ {r=0}$ which is directed along $\hat{\vec x}$ and is equal to $\hat{\vec x}w'(r)|_{r=0}$. Here the prime denotes differentiation with respect to the argument. A straightforward calculation gives
    \begin{align}\label{eq:20}
    F_{\perp x}
    &=
    n_{b0}\alpha
    K(\xi)
    ,
    \end{align}
where
    \begin{align}\label{eq:21}
    K(\xi)
    &=
    \frac{1}{2}
    \left[
    H(\xi)
    \int_0^\infty
    r'dr'
    K_1(r')
    [
    r'
    f(r')
    +
    F_1'(r')    
    ]
    \right.
    \nonumber\\
    &\left.
    -
    G(\xi)
    \int_0^\infty
    r'dr'
    K_1(r')
    (r'F_1(r')+F_2(r'))
    \right]
    ,
    \end{align}
and
    \begin{align}\label{eq:22}
    H(\xi)
    =
    \int_{-\infty}^\xi
    d\xi'
    G(\xi')
    \sin(\xi-\xi')
    .
    \end{align}

\emph{Analysis of the solution}.---Eq.~\eqref{eq:20} is our final result that proves that a transverse plasma gradient supports a perpendicular to the trajectory force in the wake of a driver bunch. The longitudinal dependence of the force is determined by functions $G(\xi)$ and $H(\xi)$ that, through Eqs.~\eqref{eq:15} and~\eqref{eq:22}, are related to the longitudinal bunch distribution $g(\xi)$. The radial beam distribution $f(r)$ also plays a role defining, together with functions $F_1(r)$ and $F_2(r)$, the integrals in~\eqref{eq:21}. 

For a numerical example, we will assume that the beam has a Gaussian distribution function,
    \begin{align}\label{eq:23}
    n_b(r,\xi)
    =
    \frac{n_{b0}}{(2\pi)^{3/2}\sigma_z\sigma_r^2}
    \exp
    \left(
    -\frac{\xi^2}{2\sigma_z^2}
    -\frac{r^2}{2\sigma_r^2}
    \right)
    ,
    \end{align}
with the rms bunch length $\sigma_z=1$ and the rms transverse size $\sigma_r=1$ ($\sigma_r = \sigma_z = k_p^{-1}$ in dimensional units). With this distribution function, $K(\xi)$ can be easily computed numerically; plot of $K(\xi)$ is shown in Fig.~\ref{fig:1}.
\begin{figure}[htb]
\centering
\includegraphics[width=0.4\textwidth, trim=0mm 0mm 0mm 0mm, clip]{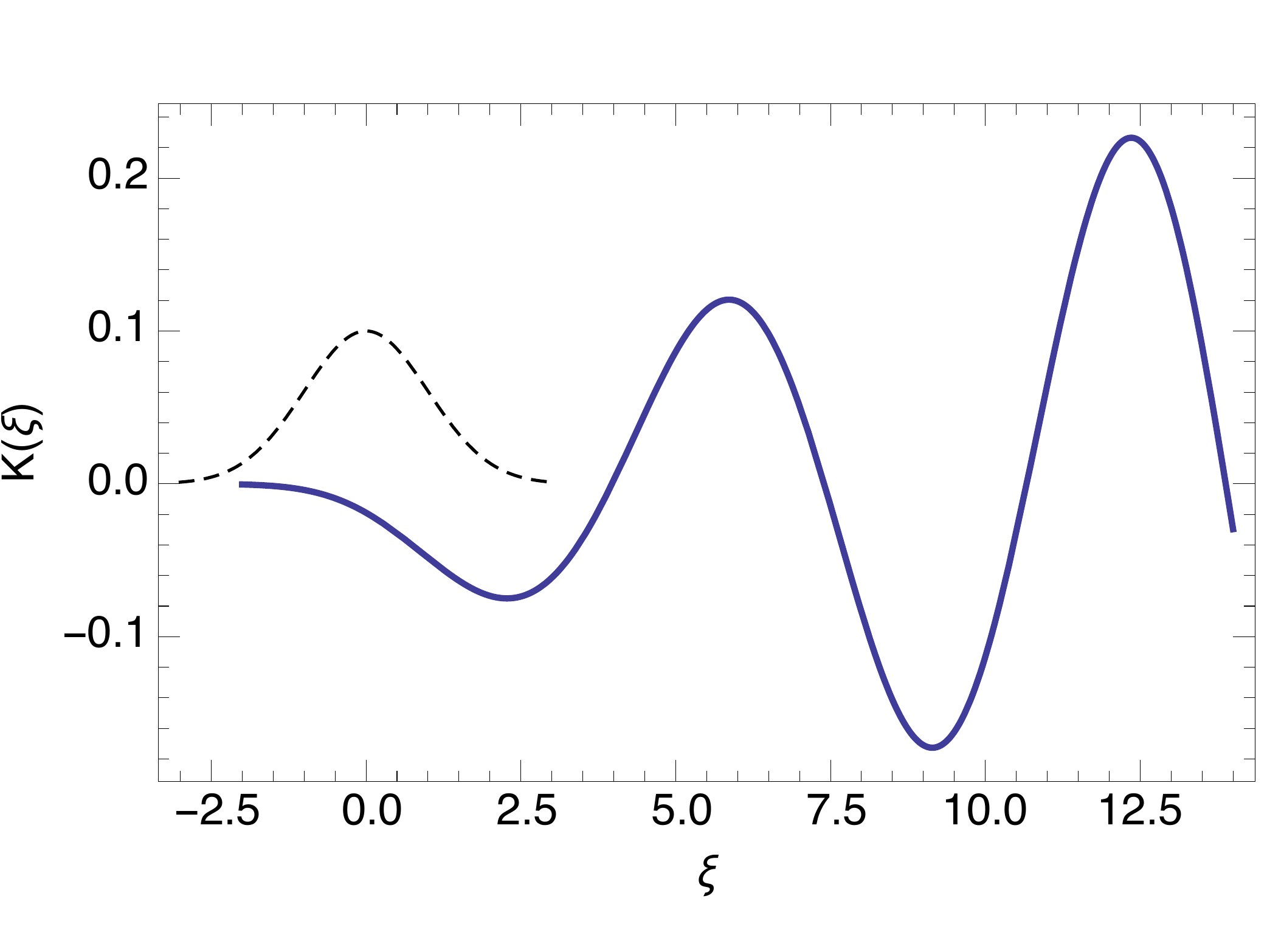}
\caption{Plot of function $K(\xi)$ (solid curve) and the Gaussian bunch distribution (dashed curve).}
\label{fig:2}
\end{figure}
The function oscillates around zero with the oscillation amplitude linearly growing with $\xi$ (these oscillations are due to the first term in the square brackets of Eq.~\eqref{eq:21} proportional to the function $H(\xi))$. The physical mechanism behind this growth is the resonant excitation of the angle-dependent density perturbation $\delta n_1$ by the axisymmetric component of the plasma waves in the wake coupled through the density gradient (see the right-hand side of Eq.~\eqref{eq:17}). In our numerical estimate we, however, will not rely on the growing values of $K(\xi)$ that it attains far behind the driver beam, and take $K\approx -0.17$ at the second minimum at $\xi\approx 9.1$ as a representative  estimate of $K$. Then the absolute value of the transverse force is given by
    \begin{align}\label{eq:24}
    |F_{\perp x}|
    \approx
    0.17 n_{b0}\alpha
    .
    \end{align}

Eq.~\eqref{eq:24} shows that the transverse force is proportional to the beam density $n_{b0}$. A large beam density, however, would lead to a nonlinear plasma flow with relativistic velocities, which is outside of the range of applicability of our linear theory that assumes $ n_{b}\ll 1$. This indicates that the maximum force is attained in the blowout regime~\cite{Kostyukov_etal,Lu_pwfa,khudik_etal:2013}, numerical analysis of which require intensive computer simulations that are beyond the scope of this paper.  To obtain a reasonable, although approximate, estimate of the force we can set $n_{b}$ and $\alpha$ in Eq.~\eqref{eq:24} to the values of the order of, but smaller than, one.  Specifically, we will choose for our estimate $\alpha=0.3$ and $n_{b0}=3$ (the latter corresponds to the maximum beam density  in Eq.~\eqref{eq:23} $n_b^{max}=0.2$) which give for the dimensionless force $F_{\perp x}=0.15$. In dimensional units, this force corresponds to the effective magnetic field $B=F_{\perp x}/e=0.15mc\omega_p/e^2$. Choosing the plasma density $n_0 = 10^{17}\ \mathrm{cm}^{-3}$ (for which $k_p^{-1}=17\ \mu\mathrm{m}$) we find $B=15.5$ T. Note that this field scales with the plasma density as $\omega_p \propto \sqrt{n_0}$, so the field will be larger in a more dense plasma.

\begin{figure}[htb]
\centering
\includegraphics[width=0.45\textwidth, trim=0mm 0mm 0mm 0mm, clip]{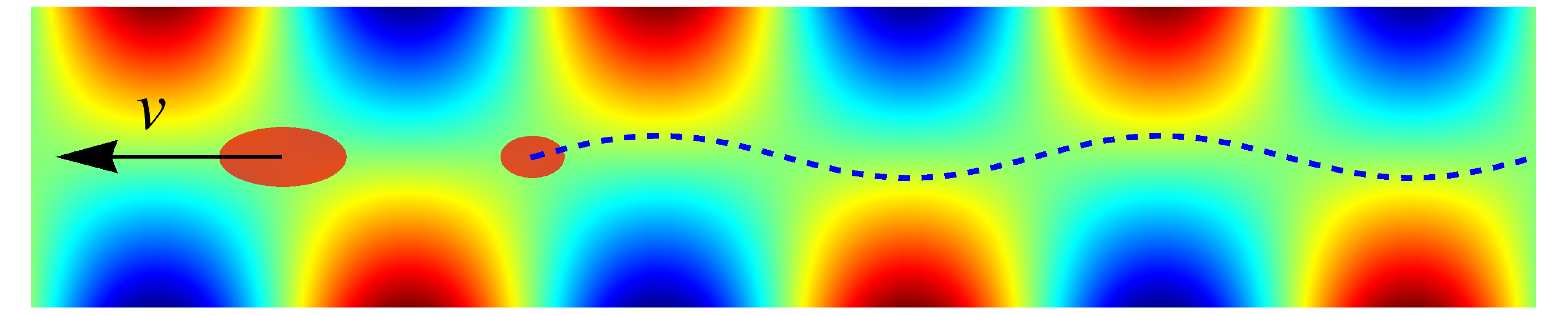}
\caption{Illustration of the undulator motion of the witness bunch in the plasma with oscillating gradient. The ambient colors indicate a periodically oscillating transverse plasma gradient.}
\label{fig:3}
\end{figure}

Our calculations assumed a constant plasma density gradient. It is clear though, that our results are also valid when $\alpha$ slowly varies along $z$ on the scale that is large compared to $k_p^{-1}$. In this case, the plasma flow calculated above will be adiabatically adjusting to the local value of $\alpha$, with the transverse force~\eqref{eq:20} becoming also a function of $z$. For a sinusoidally varying $\alpha(z) = \alpha_0\sin k_u z$, with the period $\lambda_u = 2\pi/k_u$, we obtain an undulator with linear polarization. Such an undulator field with a period of $\lambda_u = 1$ mm and the estimated above the magnetic field of 15.5 T has the undulator parameter $K =eB/k_umc^2$ equal to 1.45. A witness bunch moving in the wake of the driver is shown in Fig.~\ref{fig:3}; 
such a bunch would emit an undulator radiation in the forward direction with the wavelength $\lambda_r = \lambda_u(1+K^2/2)/2\gamma^2$, where $\gamma$ is the Lorentz factor of the witness beam. One of the important advantages of a short-period undulator is that it requires a smaller beam energy for the same $\lambda_r$. 

With a three dimensional control of the plasma gradient one can generate a helical undulator, in which the direction of the force rotates along the orbit. This is achieved by keeping the value of $\alpha$ constant and at the same time varying its direction, $\vec \alpha = \alpha_0(\hat{\vec x}\sin k_u z +\hat{\vec y}\cos k_u z)$. 

\emph{Discussion}.---Our solution for the plasma wake with a transverse plasma gradient predicts a transverse force acting on the witness bunch. An  effect of similar nature has been observed in the experiment~\cite{Muggli:2001zm,PhysRevSTAB.4.091301}, where a 28.5 GeV electron beam was reflected from the boundary of a plasma channel. In our interpretation of this experiment, such a boundary can be considered as an extreme case of the density gradient, in which the plasma radial profile can be approximated by a step function. Unfortunately, the perturbation theory developed in the paper cannot be directly applied to this experiment, because the gradient is extremely large. 

The longitudinal profile of the field in the plasma undulator varies with $\xi$, as shown in Fig.~\ref{fig:2}. This means that, depending on the position of the witness bunch and its length, particles in the head and the tail will experience somewhat different undulator fields---a property that tradition magnetic undulators do not have. This property can be used to compensate for the energy variation   in the witness bunch in longitudinal direction (an energy chirp) which typically appears in plasma acceleration.

The obtained above value for magnetic field of 15.5 T in a plasma with density $10^{17}\ \mathrm{cm}^{-3}$  is likely to be an underestimate. The present theory, while allowing to analytically calculate the transverse force, cannot be used to maximize it. As mentioned above, such an optimization requires numerical simulations in the blowout regime, where the beam density is comparable to, or exceeds, the plasma density. Judging from the results of Ref.~\cite{Lecz:2016ve}, where $10^5$ T fields were observed in simulations, although at much higher plasma density, we expect that in our problem, in an optimized regime, at least tens of Tesla magnetic fields can be achieved. Realization of short-period plasma undulators with such fields would open up an exciting prospects for all-plasma free electron lasers.

\emph{Acknowledgements}.---I would like to thank V. Khudik for help with the derivation of plasma equations.

%merlin.mbs apsrev4-1.bst 2010-07-25 4.21a (PWD, AO, DPC) hacked
%Control: key (0)
%Control: author (8) initials jnrlst
%Control: editor formatted (1) identically to author
%Control: production of article title (-1) disabled
%Control: page (0) single
%Control: year (1) truncated
%Control: production of eprint (0) enabled
%

%\bibliography{\string~/gsfiles/Bibliography/master%
%              ,THzb}
\end{document}